# Dynamic Split Tensile Strength of Basalt, Granite, Marble and Sandstone: Strain rate dependency and Fragmentation

Vivek Padmanabha[1,2] ✉, Frank Schäfer[1,2], Auriol S. P. Rae[1,3] and Thomas Kenkmann[1]

[1]Institute of Earth and Environmental Sciences – Geology, Albert-Ludwigs Universität Freiburg, Albertstrasse 23b, 79104 Freiburg, Germany.
[2]Fraunhofer-Institut für Kurzzeitdynamik - Ernst-Mach-Institute, Ernst-Zermelo-Straße 4, 79104 Freiburg, Germany.
[3]Department of Earth Sciences, University of Cambridge, Cambridge CB2 3EQ, UK.

✉Corresponding author email: vivek.padmanabha@geologie.uni-freiburg.de

## Abstract

The scope of this study is to understand the strength behaviour and fragment size of rocks during indirect, quasi-static, and dynamic tensile tests. Four rocks with different lithological characteristics namely, basalt, granite, sandstone and marble are selected. The Brazilian disc experiments are performed over a range of strain rates from $10^{-5}$ /s to $2.7 \times 10^{1}$ /s using a hydraulic loading frame and a split-Hopkinson bar. Over the range of strain rates, our measurements of dynamic strength increase are in good agreement with the universal theoretical scaling relationship of Kimberley et al. (2013). Dynamic fragmentation during a split tension mode receives very little attention and not much information is available about the generated fragment size distributions. The fragments fall into two distinct groups based on the nature of failure, coarser primary fragments and finer secondary fragments. The degree of fragmentation is assessed in terms of characteristic strain rate and is compared with existing theoretical tensile fragmentation models. The primary fragment size are less sensitive to strain rate, particularly at lower strain rates. The size of secondary fragment has a strong strain rate dependency over the entire testing range. Marble and sandstone are found to generate more pulverized secondary debris when compared to basalt and granite. Further, it is shown that the mean fragment sizes of primary and secondary fragments are well described by a power law function of strain rate.

# Highlights

- Dynamic tensile strength of the rocks are experimentally observed to obey the universal theoretical scaling relationship proposed by Kimberley et al. (2013)
- Primary (coarse) and secondary (fine) fragment's size generated from the split tensile tests are represented by a power law function of strain rate
- The mean fragment size of primary fragments are less sensitive to strain rate, while the secondary fragments have a strong strain rate dependency

# Declarations

**Funding**: This study was funded by Deutsche Forschungsgemeinschaft (project no. DFG-SCHA1612 / 2-1)

**Conflicts of interest/Competing interests**: The authors declare that there is no conflict of interest

**Availability of data and material**: The authors declare that data supporting the findings of this study are available within the article

**Code availability**: Not applicable

**Authors' contributions**: The author's contributions are listed below:
*Conceptualization*: Thomas Kenkmann, Frank Schäfer; *Methodology*: Vivek Padmanabha, Auriol S. P. Rae; *Formal analysis and investigation*: Vivek Padmanabha; *Writing - original draft preparation*: Vivek Padmanabha; *Writing - review and editing*: Vivek Padmanabha, Auriol S. P. Rae, Thomas Kenkmann, Frank Schäfer; *Funding acquisition*: Thomas Kenkmann, Frank Schäfer; *Resources*: Vivek Padmanabha; *Supervision*: Frank Schäfer, Thomas Kenkmann.

# 1 Introduction

The tensile behaviour of rocks is considered to be a vital aspect in the overall dynamic behaviour of rocks. Rocks can be subjected to dynamic loading during various events such as drilling, blasting, earthquake, landslide, and impact cratering (Kenkmann et al. 2014; Zhou et al. 2014; Zhu et al. 2015). Dynamic fracturing is a complicated process, which is dependent on mechanical properties of rock, microstructural features and the type of loading imparted. Rocks are generally weak in tension, where the uniaxial tensile strength is typically 1/10$^{th}$ of the uniaxial compressive strength of the rock (Aadnøy and Looyeh 2019; Hoek 1966). Furthermore, the dynamic tensile behaviour of rocks including fragmentation can be different to the dynamic compressive behaviour due to different strain rate dependencies. Several methods are available to characterize the dynamic tensile response of rocks at high strain rates. Among the various methods, Hopkinson pressure bar has been the most popular method for investigating the tensile dynamic behaviour of rocks, either in pure tension mode (direct tension) or Brazilian test (indirect tension) mode. Previous works related to Brazilian method of testing rocks using a split Hopkinson pressure bar (SHPB) have shown satisfactory performance (Wang et al. 2004; Zhu et al. 2015), with the advantages of easy sample preparation, simple operation and good repeatability. In 2012, the International Society for Rock Mechanics (ISRM) recommended Brazilian disc tests as an appropriate method for determining tensile strength under dynamic loading (Zhou et al. 2012). A more detailed review on dynamic tensile characterization of rocks is available in Zhang and Zhao (2014) and Heard et al. (2018).

Dynamic effects on strength are commonly represented by the 'Dynamic Increase Factor' (DIF, describing the relative enhancement of dynamic strength with respect to the static strength). Over the past few decades, extensive research into dynamic behaviour of rocks has led to the development of several DIF curves (Liu et al. 2018). These DIF curves as a function of strain rate or loading rate are generally case specific, which depend on the rock type and the nature of testing method; hence, their applicability is limited. Kimberley et al. (2013) developed a theoretical universal rate-dependent scaling relationship for the compressive strength of brittle materials incorporating micro-mechanical behaviour. The micro-crack interaction associated with the flaws and their rate dependence is explained in Paliwal and Ramesh (2008) and Kimberley and Ramesh (2011). The flaws distributed in the material play a major role in governing the strength of the material. The developed scaling relationship captures the insensitivity of strength to strain rate at low strain rates and the strong dependency at higher strain rates. Kimberley et al (2013) also showed that their scaling relationship can be used to predict the tensile strength of the material, by varying some material parameters. However, the validity of the model at high rate tensile failure is not yet proven. Li et al. (2018c) extended the Kimberley model for tensile conditions by incorporating the effects of microscale heterogeneity using Grain-Based Discrete Element Method (GB-DEM) and developed a function without altering the fundamental form. The DIF curves for granite were found to increase linearly with strain rate until intermediate strain rate and then drastically increase at higher strain rates. For a general case, Li et al. (2018a; 2018b; 2018c) recommended DIF curves in the form of a stretched Kimberley function. Additional details of the Kimberley and Li et al. model are presented later in the discussion section in conjunction with our experimental results.

During dynamic tensile failure, micro cracks (mode I) develop and grow starting from arbitrarily oriented flaws (Griffith 1921). These cracks propagate under the influence of mechanical loading and coalesce to form larger cracks. The cracks grow in size, coalesce to form multiple cracks and manifesting themselves into a network of visible fractures, leading to rock debris and fragmentation. The resulting fragment particle size vary from macro- to micro-scale. Estimation of the particle size during fragmentation can offer insights into various physical phenomena. For example: The fragment size provides important observation on the fracture mechanics of faults generated by co-seismic activity, where successive co-seismic loading leads to rock pulverisation. (Aben et al. 2016; Dor et al. 2006) . Average rock fragment size is generally used as an index in the selection and optimum usage of explosives in the mining industry (Cho and Kaneko 2004). The tensile fragments (spall) of impact craters account for significant amount of the ejecta, nearly 50% of the ejected volume was observed in the experimental work of Dufresne et al. (2013). Also, the degree of fragmentation is known to vary in different zones of an impact crater (Kenkmann et al. 2014).

Early studies on fragmentation were pioneered by Nevill Mott, who invented a theory based on an expanding cylindrical shell. The average fragment size was subsequently predicted using statistical models (Mott 1947). Since then, dynamic fragment characterization has been a subject of considerable research interests and researchers have used a variety of statistical distributions in evaluating average fragment size. Some of the common statistical distributions used are: exponential (Grady and Kipp 1985), log-normal (Ishii and Matsushita 1992), power law (Oddershede et al. 1993), Weibull (Brown and Wohletz 1995) and others (Ouchterlony 2005; Sil'vestrov 2004). Another group of researchers have developed models based on principles of energy balance (Glenn and Chudnovsky 1986; Grady 1982; Yew and Taylor 1994). According to energy based fragment size models, the fragment size is governed by the balance between externally imparted energy and the internally developed energy on the surfaces of the fragments. Several numerical models were also developed in order to include the effect of stress waves (Drugan 2001; Levy and Molinari 2010; Miller et al. 1999; Zhou et al. 2006). The above listed theoretical and computational models are generally considered for the case of uniaxial tension stress state.

Experimental studies of dynamic tensile fracturing are commonly carried out using spallation techniques (Grady 1988; Grady 2006). Split Hopkinson Tension Bar (SHTB) are proven to be a reliable test facilities to dynamically characterize the fragments under tensile loading. Griffith et al. (2018) used SHPB facility to generate tensile radial stress in the rock sample using expanding cylinder theory (Shockey et al. 1974). Their experiments suggested that the fragmentation process have a strong strain rate dependency and the transition from fragments to pulverization occurs at a strain rate in the order of $10^2$ /s.

The dynamic split tensile test (Brazilian test) is generally not considered favourable for fragmentation studies, as the indirect tension test initiates and propagates a single fracture. Such fracture behaviour is often observed during quasi-static loading. At higher strain rates, a complex stress interaction takes place within the sample leading to multiple fragments and the mass percentage of the fragments were found to increase (Zhu et al. 2020). It is important to quantify the fragments generated from such complex stress conditions. Fragments resulting from the dynamic split tension tests are generally of two different sizes: coarse sized fragments (mostly of semi-disc type) from the primary fractures and finer debris from secondary fractures (Cai 2013). The secondary fractures play a major role in the dynamic fragmentation process, which is often overlooked in the fragment analysis. Very little information is available in the existing literature (Li et al. 2018a; Zhu et al. 2020) on the dynamic fragmentation of dynamic Brazilian tests and there is no data concerning the size distribution of fragments.

Therefore, there is a need for an in-depth analysis and characterization of dynamic strength and fragmentation in split tensile test mode.

In this study, using dynamic Brazilian disc testing, the tensile strength of rocks of different lithologies is investigated using a SHPB at intermediate strain rate range ($10^0$ /s - $10^2$ /s). We discuss the DIF associated with strain rate, and the applicability of the universal theoretical scaling relationship of strength. Additionally, fragment size distributions of the experimental products (primary and secondary fragments) are measured and the strain rate dependency of the fragment sizes are systematically quantified. Finally, the experimental results are compared with the existing theoretical models on the dynamic fragmentation and the acceptability of such models for split tensile fragments are discussed.

## 2 Experimental details

### 2.1 SHPB test facility and principles

The dynamic split tensile tests were carried out using a split Hopkinson pressure bar (SHPB) facility at the Geology Department, Albert-Ludwigs Universität Freiburg, Germany. The SHPB consists of three 50 mm diameter bars, each made of Titanium alloy ($E_B$ = 110 GPa, $\rho_B$ = 4.43 g/cc), these are termed the striker bar, incident bar, and transmission bar. A striker bar of length 250 mm is housed inside a barrel connected to a pressure vessel. To avoid wave reflection during test time, length of the incident and transmitted bars were designed to be 2500 mm. The end of the transmitted bar is made to pass through a momentum trap system, where motion of the bar is arrested. In order to achieve 'dynamic force equilibrium', it is necessary to use a pulse shaper between the striker and incident bar. This results in a slowly rising incident pulse and avoids wave dispersion effects in brittle materials (Frew et al. 2002; Zhang and Zhao 2014). In our study, we have used aluminium foam of 10 mm thickness and 90% porosity as a pulse-shaper. To achieve ideal pulse shapes for the experiments in this study, the aluminium foam was pre-hit at a striker velocity of ~10 m/s resulting in a final thickness of ~7.5 mm (Rae et al. 2020; Zwiessler et al. 2017). **Fig. 1**. Illustrates the schematic diagram of the SHPB bar.

The cylindrical rock sample is placed diametrically between the incident and transmitted bar. The compressed gas released from the pressure vessel accelerates the striker bar, which in turn strikes the incident bar via the pulse shaper. A compressive elastic wave generated in the incident bar travels towards the rock sample. Due to the change in material impedance at the bar-sample interface, part of the compressive wave is reflected, while, the remaining part of the wave is transmitted through the sample into the transmission bar. During this process, the sample must be uniformly compressed and undergo homogeneous deformation in compression experiment. However, in the case of Brazilian test, the sample undergo spatially non-uniform stress distribution. But the forces at the ends of the bars should be equal for the experiment to remain in 'dynamic force equilibrium' condition. In addition to dynamic force equilibrium, Brazilian disc tests require that the crack initiates at the centre of the specimen where the sample is under tension.

During the dynamic Brazilian tests, the compressional waves generated from the incident bar transmits radially into the cylindrical sample. Waves with higher incidence angles reflect at the circular free surface of the sample as a tensile stress pulse reaching the diametrical line of the sample (For detailed derivation, refer Zhou et al. (2014)). Zhou et al. (2014) observed the stress pulses with α = 30° (α is the incident angle over which the waves are radially distributed, see **Fig.1**), are distributed radially to reach the centre of the sample earliest and thence the centre becomes the most vulnerable point in the sample for a tensile failure.

The response of the test sample is determined using wave propagation theory (Kolsky 1963). The axial stress waves induced in the incident and the transmission bars are recorded using strain gauges mounted on the respective bars, consequently, three strain measurements were made: (i) incident, $\varepsilon_i$, (ii) reflected, $\varepsilon_r$ and (iii) transmitted, $\varepsilon_t$. A digital oscilloscope records the voltage signals at a sampling rate of 1.25 MHz. The noise in the strain signals are filtered and Pochhammer-Chree dispersion correction (see Chen and Song (2011) and Rigby et al. (2018) for further details) is applied thereafter. The force accumulated on the incident ($F_1$) and transmitted ($F_2$) bar ends are evaluated using **Eq. 1** and **Eq. 2:**

$$F_1 = E_B A_B [\varepsilon_i(t) + \varepsilon_r(t)] \quad (1)$$

$$F_2 = E_B A_B [\varepsilon_t(t)] \quad (2)$$

$$A_B = \frac{\pi D_B^2}{4}$$

, where, $A_B$ is the cross sectional area of the bar, $E_B$ is the elastic modulus of the pressure bar and $D_B$ is the diameter of the SHPB bar.

For the test sample is in the state of dynamic force equilibrium, we have:

$$F_1 = F_2 \quad (3)$$

The dynamic split tensile strength of the rock samples can be determined using either the peak load generated on the incident end or transmitted end of the sample (Jin et al. 2017). Ideally with the assumption of force equilibrium, both the values should yield same tensile strength values. A perfect dynamic equilibrium is not always possible and considering the experimental errors, an average value between them is considered to be the most accurate result. Dynamic tensile stress, $\sigma_t(t)$:

$$\sigma_t(t) = \frac{(2F_1)}{\pi D_S T} \text{ or } \frac{(2F_2)}{\pi D_S T} \Rightarrow \frac{2E_B A_B}{\pi D_S T}[\varepsilon_i(t) + \varepsilon_r(t)] \text{ or } \frac{2E_B A_B}{\pi D_S T}[\varepsilon_t(t)]$$

$$\sigma_t(t)_{avg.} = \frac{(F_1 + F_2)}{\pi D_S T} = \frac{E_B A_B}{\pi D_S T}[\varepsilon_i(t) + \varepsilon_r(t) + \varepsilon_t(t)] \quad (4)$$

, where, $D_S$ and T are the diameter and thickness of the cylindrical disc of rock sample.

## 2.2 Rock samples and sample preparation

In the present study, we investigate four different types of rocks of igneous, sedimentary and metamorphic origin. Samples of basalt, granite (igneous), sandstone (sedimentary) and marble (metamorphic) with densities of 2.90, 2.62, 2.04 and 2.70 g/cm$^3$ respectively, were collected from different lithostratigraphic units : fine-grained basalt was collected from Hegau, Germany (referred hereafter as 'HeBa'); pale pink, coarse-grained granite was collected from Malsburg, Germany (MaGr); fine-grained, porous sandstone was collected from Seeberg, Germany (SeSa); and lastly, calcite dominated marble was acquired from Carrara, Italy (CaMa). Quasi-static mechanical properties of the rocks were carried out using a FORM+TEST Alpha 2-3000 hydraulic loading frame. With a minimum of three samples per rock type, stress controlled quasi-static Brazilian tests were performed with loading rates from 0.05 to 0.15 kN/s. The physical and mechanical properties of the rocks used in the present study are summarised in **Table 1.**

The Brazilian disc samples were prepared according to the recommended ISRM standards (Zhou et al. 2012) for SHPB testing. Uniform, representative, cylindrical samples of diameter 41 ± 0.25 mm were drill-cored from large blocks of each lithology. According to the ISRM recommendation, smaller diameter/harder samples should be prepared with a 1:1 slenderness ratio and larger diameter/softer samples should be prepared with a slenderness ratio of 0.5:1 (Mishra et al. 2020). In the present study, all four types of rocks were prepared with two sets of length to diameter ratio, 0.5:1 and 1:1. The diametrical surfaces of the sample were made flat, such that the surfaces are perpendicular with the loading axis. A total of 40 cylindrical samples were prepared: 10 nos. of HeBa sample, 10 nos. of MaGr Samples, 12 nos. of SeSa and 8 nos. of CaMa samples. The samples were labelled after their rock type in a sequential order.

Additional modifications were made to the cylindrical samples to facilitate the dynamic force equilibrium and centrally initiated crack conditions. To prevent compressive stress concentration and failure at the loading ends (between the sample and the bar), cylindrical samples are recommended to have a flattened end (Rodríguez et al. 1994; Wang et al. 2004; Wang et al. 2006; Wang et al. 2009). The two cylindrical faces of the samples in contact with the bars were trimmed and flattened, such that the flat ends are parallel to each other. The loading from the bar onto the sample is thus distributed over the flattened area. The width of the flat portion is governed by the loading angle, 2α (shown in **Fig. 1**). In the theoretical and experimental studies of Wang et al. (2004), 2α = 20° was found to guarantee a central crack initiation. All the samples in the present study were flattened as per that recommendation. Furthermore, based on the Griffiths strength criteria, the tensile stress of the flattened Brazilian disc is modified for 2α = 20°. The final expression is (Wang et al. 2006):

$$\sigma_t(t)_{avg.} = \frac{0.95(F_1 + F_2)}{\pi D_s T} \tag{5}$$

## 2.3 Analysis and Data processing

### 2.3.1  Force equilibrium and validation

As mentioned in the previous section, the prerequisites for SHPB testing of Brazilian disc are the 'dynamic force equilibrium' and 'central crack initiation' in the sample. The signals recorded by the strain gauges on the incident and transmitted bar are processed and the forces developed at the end of the bars are evaluated using **Eq. 1** and **Eq. 2**. **Fig. 2a** shows typical incident and reflected signals, with the corresponding forces generated at the bar ends. The forces at each end of the sample remain approximately equal throughout the duration of the experiment. This indicates that the dynamic force equilibrium is achieved and the sample remained in the state of equilibrium before failure.

Furthermore, it is important to ensure that the crack originates at the centre of the sample. Generally, a high-speed camera can be used to monitor the crack propagation and the subsequent fracture process (Jin et al. 2017; Li et al. 2018b). Alternatively, multiple strain gauges can be placed on samples for the same purpose (Wang et al. 2016; Zhou et al. 2014), however, the sample dimensions in this study are too small to mount multiple strain gauges. Instead, a simpler and more cost effective method, based on electric potential drop was employed. Interconnected electric circuits in the form of grids are painted on the surface of the rock sample using electrically conductive paint (Bare Conductive, London, UK). Two such mirror image circuits are marked on the incident (left) and the transmitted (right) ends of the rock sample, as shown in **Fig. 2b** (inset figure). A Wheatstone bridge balances the two legs of the circuit. The circuits are activated by passing a constant current through them and the electric potential across the circuits are continuously monitored via a signal amplifier. **Fig. 2b** shows the voltage–time signal recorded from the left and right paint circuit. It was found that the two signals from the circuits were active over the same time interval with coordinated start and ending time, and each signal shows three step-wise decreases in the voltage with increasing time. However, the sectional voltage amplitudes at the three locations are different. Overall, this suggests that the propagating cracks have broken the three grids travelling at variable velocities, in the opposite direction. It also implies that the crack has originated close to the centre of the sample and therefore, that the sample did fail in tension.

### 2.3.2  Determination of strain rate

Measurement of strain rate during the deformation is an important aspect of the dynamic testing. During a traditional compressional SHPB testing, the strain rate is normally calculated from the strain signals measured on the bar or an approximate value is deduced from the velocity of the striker bar and length of the sample (Rae et al. 2020; Shin and Kim 2019). Because of the non-uniform stress state in Brazilian disk sample, both methods will not yield a representative tensile strain rate. Thus, in this study, an additional strain gauge was placed on the sample surface to allow determination of the strain rate up to the point of failure. In all our test samples, a strain gauge (HBM, 1-LY66-6/120) was mounted on the centre of the rock sample surface using a HBM X60 adhesive, such that the loading axis is perpendicular to the gauge axis, i.e., in an orientation, where the strain gauge measures the tensile strain. A schematic diagram of the strain gauge mounted flattened Brazilian sample is shown in **Fig. 1**.

A typical strain gauge signal recorded from MaGr02 is shown in **Fig. 3a**, the strain signal values are normalized (between 0 and 1) for comparison with the calculated tensile stress. The strain remains at zero, until the stress signal experiences a sudden rise. At this point, the strain begins to rise gradually before abruptly increasing and the signal being cut-off. The abrupt increase of strain indicates that the fracture is growing in the sample. **Fig. 3b** shows this stage of failure during the time interval from 0.1 to 0.275 ms. The failure initiation can be more clearly identified using the first derivative of strain signals, shown in **Fig. 3b**. The start of the material deformation is the point, where the initial perturbation happens in the strain rate signal history and the end of the failure is when the '$\varepsilon$' signal shows an abrupt increase (Griffith et al. 2018). The strain rate is determined by taking the slope of the strain curve over this macroscopic failure period (from point A-B in **Fig.3b**). **Fig. 4** shows representative plots of tensile stress and strain history from each of the four different rock types (the region over which strain rate is determined is highlighted in grey colour band). In all the test cases, the end of the failure zone is observed in the close vicinity of peak stress.

## 3 Experimental Results and Discussion

### 3.1 Dynamic split tensile strength and its strain rate dependency

Based on the methods described in section 2.3, the tensile strength of the rock samples and the strain rate of each experiment were evaluated. **Table 2** lists the values of strain rates and corresponding split tensile strength values for all the test cases; the experimental uncertainty of stress and strain rate are expressed as errors. In the present experimental series, the quasi-static strain rates ranged from $10^{-5}$ /s to $10^{-3}$ /s and the strain rates achieved by the SHPB experiments ranged from $4 \times 10^{-1}$ /s to $2.7 \times 10^{1}$ /s. **Fig. 5** shows the variation of tensile strength with the strain rate under quasi-static and dynamic conditions for all lithologies. Overall, the dynamic tensile strength of the rocks is higher than the quasi-static tensile strength (1.5 to 5 times) and there is a strong dependency of tensile strength on strain rate. The increase in the strength behaviour can be explained from a microstructural viewpoint, where the micro cracks plays a crucial role in rock failure (Chen et al. 2018). The nature of developed micro cracks depend on both external loading and inherent material fabrics and/or pre-existing flaws. During rapid high strain rate loading, the weakest micro flaw lags the increment loading for the crack to grow. Thus additional strong flaws in the material gets activated to accommodate the applied strain (Ramesh et al. 2015). Before the macroscopic failure occurs, increasingly strong flaws are activated and more such flaws fractures with individual fragments are generated.

The strain-rate dependency of the dynamic split tensile strength of the four different rocks are shown in **Fig. 5**. In absolute terms, quasi static strength of the rocks are highest for basalt, followed by granite, marble and sandstone. The dynamic strength are observed to proportionately increase in the same order. Among the four chosen rocks, Seeberger Sandstone is highly porous (about 23%, Poelchau et al. (2013)) and permeable, where grains are weakly bonded with a silicate cement (Kenkmann et al. 2011). Porosity in rocks are principal sources of micro flaws (Kranz 1983; Wong et al. 2004). The evolution of micro-cracks, in a porous sandstone is predominant from that of non-porous rocks like basalt, granite or marble. Studies have shown that micro cracks originating from the microscopic flaws significantly influence the dynamic strength of the material (Daphalapurkar et al. 2011). In addition to micro cracks (Huang et al. 2002), heterogeneity of rocks also play an important role in the increase of the dynamic tensile strength (Cho et al. (2003).

The increase in the dynamic tensile strength can be better understood using DIF, the dynamic strength normalized by the quasi-static strength of the material. Generally, power laws are used to fit the DIF ($\sigma_t/\sigma_o$) as a function of strain rate or loading rate (Doan and d'Hour 2012; Grady and Lipkin 1980; Lankford 1981). However, Kimberley et al. (2013) developed a universal rate-dependent theoretical scaling relationship incorporating the material's microstructural properties. The interaction of pre-existing flaws and dynamics of the micro crack growth have shown to be important parameters in describing the strength of the brittle materials. The model describes characteristic strength ($\sigma_o$) and characteristic strain rate ($\dot{\varepsilon}_o$) by incorporating mechanical (Young's modulus (E), fracture toughness ($K_{IC}$), limiting crack speed ($c_d$)) and microstructural (flaw size ($\bar{s}$), flaw density ($\eta$)) parameters. The functional form of characteristic stress and characteristic strain rate is shown in **Eq. 6**, as described in the original work of Kimberley et al. (2013).

$$\sigma_o = \alpha \frac{K_{IC}}{\bar{s}\eta^{1/4}} \; ; \; \dot{\varepsilon}_o = \alpha \frac{c_d K_{IC} \eta^{1/4}}{\bar{s}E} \qquad (6)$$

The characteristic stress is related to the stress required to generate a crack such that the inherent flaws in the material can be bridged together; the parameter, α, ensures that the value of $\sigma_0$ corresponds to the strength of the material. The characteristic strain rate is the critical strain rate at which the strength of the rock is double the quasi-static strength (DIF = 2). The universal theoretical scaling relationship in terms of characteristic strength and characteristic strain rate is shown in **Eq. 7** (Kimberley et al. 2013) :

$$\frac{\sigma_t}{\sigma_o} = 1 + \left(\frac{\dot{\varepsilon}}{\dot{\varepsilon}_o}\right)^{2/3} \qquad (7)$$

Kimberley et al (2013) have stated that their theoretical model predicts well the behaviour of brittle materials (ceramics and geological materials) at both compression and tensile conditions. With regard to the compressional behaviour their model has been verified, but very limited data were available in tension to make a detailed assessment. Hogan et al. (2015), explored Kimberley relation in tension condition by fitting their indirect tension experimental data (using Brazilian disc technique) on meteorite samples at low strain rates.

Li et al. (2018c) questioned the validity of the Kimberley model, in particular to tension loading. Li et al. (2018c) developed a model similar to Kimberley model but treated the exponent as a free parameter. Their model was based on numerical simulation and recommended a more fundamental form (shown in **Eq. 8**) for DIF, the proposed equation can be stretched with the increase rate parameter (β), taking any positive integer value. The stretched function makes it feasible to have an applicability over a wide range of higher strain rates. A review of the experimental data in dynamic tension (direct and indirect) along with the regression results is presented in Li et al. (2018a), the β factor varied from 0.35 to 0.63.

$$\frac{\sigma_t}{\sigma_o} = 1 + \left(\frac{\dot{\varepsilon}}{\dot{\varepsilon}_o}\right)^{\beta} \qquad (8)$$

In the present experimental series, the characteristic stress and characteristic strain rate for individual rocks values are obtained by nonlinear least-square fitting (**Eq. 8**) to the experimental data set of each rock. The regression values of the dependent variable along with characteristic values are shown in the table in **Fig. 5**. For the rocks under investigation in Brazilian tests, β is found to vary from 0.54 to 0.71. With β being a free parameter, the characteristic strain rate ($\dot{\varepsilon}_o$) of the investigated rocks in tension are determined to be: Basalt = 1.32 ± 0.96 /s; Granite = 2.09 ± 1.63 /s; Sandstone: 1.49 ± 0.68 /s; Marble = 2.83 ± 2.50 /s. And the characteristic stress ($\sigma_o$) of the investigated rocks are: Basalt = 15.03 ± 2.56 /s; Granite = 8.16 ± 2.00 /s; Sandstone: 4.38 ± 0.60 /s; Marble = 6.72 ± 1.85 /s. These characteristic values indicate the relative dynamic strength of the investigated rocks compared to their quasi-static strength. Further, the flaw density and flaw size for the particular rock type can be technically determined using the characteristic values in **Eq. 6.**

The experimentally observed results are graphically compared in the normalized form with the theoretical model of Kimberley in **Fig. 6**. The tensile strength and strain rate listed in **Table 2** are normalized against their respective rocks characteristic values. Considering the experimental uncertainty, with β value of 0.583 ± 0.012 for the global fit (within 2 standard deviation errors), the fitted curve is considered to be in good agreement with the Kimberley model. The present experimental study further establishes the performance of universal rate-dependent model without β being a free parameter, but with a fixed value of 2/3 in tension mode. The curve fitting procedure has been repeated again with β = 2/3, to determine the definitive characteristic strain rate for the rocks under investigation. The revised characteristic strain rate values for the rocks are: Basalt = 2.40 ± 0.68; Granite = 2.52 ± 1.01; Sandstone: 2.61 ± 0.56; Marble = 2.39 ± 1.15. The $\dot{\varepsilon}_o$ values shows that basalt and marble are more sensitive to strain rate, followed closely by granite, and then by sandstone.

Rae et al. (2020) found that the Kimberley model to be in good agreement for felsic crystalline rocks in compression. The characteristic strain rate of Malsburg Granite (examined in the present study) in compression was found to be 217 ± 95 /s (Rae et al. 2020). In another study, the characteristic strain rate of Seeberger Sandstone and Carrara Marble were reported to be 170 /s and 65 /s respectively (Zwiessler et al. 2017). The characteristic strain rate in tension is about 1 to 2 orders of magnitude lower than the compressive characteristic strain rates. The ratio between compressive and tension characteristic strain rate values could be lithological dependent, which remains to be investigated; it is considered beyond the scope of this article.

**3.2 Dynamic Fragmentation**

A typical dynamic Brazilian test performed using SHPB will result in four different types of fragments (Zhu et al. 2020), namely, Type I – semi-disc, Type II – section fragments, Type III – small sized debris and Type IV – powder. Type I and Type II are coarse sized fragments which are primarily caused due to lateral tension failure. Type I fragments are generally two large sized semi-circular disc shaped fragments. Type II fragments are flake like split fragments emerging from the tensile failure. Type III fragments are small sized section fragments due to shear failure, generally appear close to the bar ends (Dai et al. 2010), Type IV fragments are mostly in the pulverized state, generated around the shear and tensile fracture surfaces. In the present study, Type I and Type II fragments are categorised as coarse fragments (primary) and they are mainly bounded by tensile fractures (mode I); Type III & IV fragments are finer particle fragments (secondary) resulting from different kind of failure modes, to a greater extent by shear failure. Therefore, secondary finer fragments cannot, in themselves, be classified under specific failure modes. The fragment morphology of different rocks (HeBa, MaGr, SeSa, and CaMa) at different strain rates with the four fragment types are highlighted in **Fig. 7**.

Particle size distributions were measured for all the fragmented samples collected after failure using sieves. Standard sieves with square apertures of 16, 6.3, 2, 1, 0.63, 0.4 and 0.2 mm were used and particles finer than 0.2 mm were collected in a pan. Several distribution functions have been used to fit the size distribution of the fragments generated from high dynamic events namely power law, lognormal, Weibull, Gilvarry, Swebrec; the most popular being Weibull distribution for impact fragmentation (Cheong et al. 2004). **Fig. 8** presents the Fragment Size Distribution (FSD) data and fitted cumulative Weibull distributions for basalt, granite, sandstone and marble at different strain rates. The goodness-of-fit is largely considered to be extremely good for all the test cases, except for few test cases of granite at higher strain rates (24.42 /s and 27.14 /s). In the FSD's, the weight of fragments retained on each of the sieves has been expressed as the percentage of the total weight of the sample and subsequently, the cumulative weight of the fragments smaller than size 'D', $P(<D)$ is determined. For all the test cases, the passing weight percentage of the fragments increases with strain rate at all particle sizes.

The inset bar graphs in **Fig. 8** show that the largest sieve retains more than 60 % of the fragment mass and the retained percentage mass decreases with increase in the strain rate. Since majority of the Type I and Type II fragments are collected in either 6.3 mm or 16 mm sieve; particles retained on 6.3 mm or higher aperture sieve, are segregated as primary fragmented particles and the rest of the particles passing through 6.3 mm sieve as secondary fragmented particles.

ISO standards (ISO 9276-3:2008) recommend Rosin-Rammler (Weibull) distribution and Gates-Gaudin-Schuhmann (bilogarthimic) distribution for the extreme value analysis of the coarse and fine particles respectively. Sanchidrián et al. (2014) have performed a detailed analysis on the high strain deformed rocks and they recommended Grady, Weibull and Swebrec functions as an ideal choice, when $P(<D)$ lies between 20% and 80% passing (coarse fragments). For fine fragments with below 20% passing, bi-component distribution like bi-modal Weibull and Grady are preferred. In the present study, for a particle/fragment size of 6.3 mm, the cumulative weight are found to be well below 20 % passing. Which further suggest that, the coarse and fine particle fragments can be classified with reference to the sieve size 6.3 mm.

### 3.2.1 Measurement of primary fragments

The primary fracture fragments of the rocks splitting into two half-disc geometries (Type-I) and angular flaky fragments along the loading direction (Type II) are shown in **Fig. 7**, under primary fragments. At low strain rate conditions, the cylindrical sample generally splits into two halves and as the strain rate increases, the discs are severely damaged (resulting in fractural debris). A cumulative fragment size distribution for each of the rock type is fitted to the sieve analysis data using the two-parameter Weibull distribution. The cumulative density function of Weibull distribution is expressed as:

$$P(<D) = 1 - exp\left[-\frac{D}{S_o^p}\right]^{n_p} \tag{9}$$

, where, P(<D) is the cumulative weight percent of all the fragments smaller than particle size (D); $n_p$ and $S_o^p$ are fitting parameters. The parameter '$S_o^p$' is scale factor, interpreted as a characteristic dimension of the fragments or maximum diameter (Wu et al. 2009) of the fragments over the accumulated range. The parameter '$n_p$' is the shape factor, which represents the range of fragment size distribution; it is also referred to as the Weibull modulus (or uniformity index). The Weibull parameters are derived from the experimental sieve data shown in **Fig. 8**. As the distribution is mostly dominated by Type I and Type II fragments, the characteristic size ($S_o^p$) and uniformity index ($n_p$) of the distribution represents the features of coarse sized primary fragments.

The primary characteristic fragment size ($S_o^p$) is plotted as a function of strain rate for rocks HeBa, MaGr, SeSa and CaMa in **Fig. 9i a-d**. For a comprehensive understanding, the characteristic size of the fragments for each of the rocks are plotted along with the characteristic values of the theoretical models derived for the respective rock type. For comparison with the experimental data, the following average $K_{IC}$ values are chosen for the theoretical model (Atkinson and Meredith 1987) : $K_{IC\_Basalt}$ = 2.58 MPa m$^{0.5}$, $K_{IC\_Granite}$ = 1.73 MPa m$^{0.5}$, $K_{IC\_sandstone}$ = 0.9 MPa m$^{0.5}$ and $K_{IC\_marble}$ = 1.16 MPa m$^{0.5}$. A review of the existing theoretical models includes: Grady model (Grady 1982); GC model (Glenn and Chudnovsky 1986); YT model (Yew and Taylor 1994); Zhou et al. model (Zhou et al. 2006); YTGC model (Jan Stránský 2010). The expression for characteristic fragment size proposed by the above mentioned models are summarized in Li et al. (2018a). Among the various fragmentation models, Grady, GC and Zhou et al. models appear to fit best to the presented data. **Fig. 9i** shows the characteristic size of fragments from the experiments are bounded between Grady model and YTGC model. The Grady model has been considered to overestimate the characteristic size, particularly at lower strain rate (Griffith et al. 2018). In the present study, characteristic values at lower strain rates (1- 10/s) show no significant difference in the characteristic values, but slightly decrease as the strain rate increases,. Such a behaviour is described in the GC model; however, the GC model tends to over predict the present experimental results. At intermediate strain rates (10 - 27/s), the measured values are more closely matched by the Zhou et al model than the GC model, except for the porous SeSa. The characteristic dimension of the SeSa are much lower than the Zhou et al model predicts. As discussed earlier, the sandstone rock is highly porous and crack branching process is quite active from the other three rock types. Even at low impact experiments, the dominant fragments of sandstone were observed to be barely intact, which indicates the rock has undergone an early shear failure fracture, at lower strain rate.

The shape factor or uniformity index ($n_p$) represents the homogeneity of the fragment size distribution, a higher value corresponds to a homogeneous set with a uniform fragment size, whereas a lower value represent heterogeneous set with a wide distribution of fragment size (Lu et al. 2008). The influence of strain rate on the uniformity index ($n_p$) is shown as a scatter plot in **Fig. 9 ii.** The $n_p$ value of the fragment size distribution, is found to be rate dependent and decreases with increase in strain rate. The trend of the index values with respect to the strain rate, suggests that beyond a transitional strain rate (between 10 /s - 20 /s), the index value remains constant over a small bandwidth. Interestingly, around the zone of this transitional strain rate, the characteristic size value starts to decrease (as seen in **Fig.9 i**). The transitional strain rate for sandstone (SeSa) could be much less than 10 /s. Additional experimental data are required beyond transitional strain rates for further understanding.

Unfortunately, with the present experimental setup it is difficult to attain high strain rates in the Brazilian test mode.

The statistical properties of Weibull distribution for primary (coarse) fragments are also derived using the formula: (i) mean, $\mu_{\text{p-mean}} = S_o^p\, \Gamma(1 + 1/n_p)$ and (ii) variance, $\sigma_p^2 = S_o^{p^2}\Gamma(1 + 2/n_p) - \mu_{p-mean}^2$, where, $\Gamma$ is the gamma function. The mean of the fitted Weibull cdf is interpreted as the 'Mean particle size, $\mu_{\text{p-mean}}$' of the primary fragments, which are moderately lower than the characteristic size values.

### 3.2.2 Measurement of secondary fragments

The secondary fragments involve complex fracture processes with different kinds of failure modes, mostly dominated by the shear cracks originating from pre-existing flaws. These shear cracks will be accelerated under dynamic conditions, leading to fine fragments (Momber 2000). In the previous section, for coarse-grained particle fragments, Weibull distribution cdf was found to well represent the experimental data. However, if the analysis is focused on the finer portion of the fragments, i.e. when the fragments size is very small when compared to characteristic size (D << $S_o$), Weibull cdf (**Eq. 9)** gets reduced to (Momber 2000; Turcotte 1986; Wu et al. 2009) the form shown in **Eq. 10**. Where, S(<D) is cumulative weight percent of all the fine fragments and $S_o^s$ and $n_s$ are shape and scale factors respectively for the secondary fragments.

$$S(<D)\% = \left(\frac{D}{S_o^{s*}}\right)^{n_s} \tag{10}$$

It is interesting to observe that the reduced form of Weibull cdf distribution is similar to the Gates-Gaudin-Schuhmann distribution (Macías-García et al. 2004; Turcotte 1986). **Eq. 10** is further transformed into a linearized function by applying natural logarithm, which yields:

$$ln\,\frac{S(<D)}{100} = ln\left(-\frac{D}{S_o^s}\right)^{n_s} \tag{11}$$

$$ln\frac{S(<D)}{100} = n_s \cdot ln\,D - n_s \cdot ln\,S_o^s \tag{12}$$

**Eq. 12**, is in the linear form y = m (x) + C, which can be graphically represented with ln (S<D)/100 as the y-axis and ln (D) as the x-axis. The slope of the linear fit data gives us the shape factor, '$n_s$' and the characteristic size for secondary fragments, $S_o^s$ is obtained from the y-intercept. It is important to note that, S(<D) is the cumulative weight percent of all secondary fragments which are passing through 6.3 mm and retained on 2 mm and below sieve sizes, viz. the primary fragments are removed in the analysis. The graphical natural log-log plot of secondary fragments for basalt, granite, sandstone and marble rocks are shown in **Fig.10**. The individually derived parameters of the distribution at varying strain rates are mentioned in **Fig.10** (inset-table), the coefficient of determination ($R^2$) values are found to be greater than 0.970. When compared to primary fragments, the uniformity index ($n_s$) value does not vary much with increase in the strain rate, meaning the distributions have a similar D-value (also called the fractal dimension, D = 3 - $n_s$). The average D-values for the basalt, granite, sandstone, and marble is 2.103, 2.239, 2.829, and 2.730 respectively. This indicates that the fragment size distributions are self-similar.

Similar to primary fragments, the statistical properties of Gates-Gaudin-Schuhmann distribution for secondary (fine) fragments are evaluated using: (i) mean, $\mu_{s\text{-mean}} = (S_o^p \, n_s) / (1 + n_s)$ and (ii) variance, $\sigma_s^2 = S_o^{p\,2} \, [n_s/(n_s+2) - n_s^2/(n_s + 1)^2 \,]$.

### 3.2.3 Normalization of fragment size

Dynamic fragmentation of rocks is commonly treated as a statistical process, which directly depends on many inherent rock properties (density, modulus, mineralogical composition, microstructural features etc.) and mechanical loading parameters (strain rate, testing method). It would be convenient to represent the fragmentation products in a dimensionless quantity using relevant normalization parameters. In this section, the strain rate ($\dot{\varepsilon}$) and the mean fragment size ($\mu_{p\text{-mean}}$ and $\mu_{s\text{-mean}}$) are normalized over characteristic strain rate ($\dot{\varepsilon}_o$) and characteristic length ($L_o$) respectively. The characteristic length, $L_o$, is the characteristic length scale of the system. In terms of theoretical modelling, it is the distance travelled by the stress waves over the characteristic time ($t_o$), delivered within cohesive element, which is defined by Camacho and Ortiz (1996):

$$t_o = \frac{K_{IC}^2}{c_p \sigma_t^2} \tag{13}$$

$$L_o = c_p \cdot t_o \rightarrow \frac{K_{IC}^2}{\sigma_t^2} \tag{14}$$

, where, $\sigma_t$ is the quasi-static tensile strength and $c_p$ is the P-wave velocity of the rock. The reference values of $K_{IC}$ used in **Eq. 14** are mentioned in Section 3.2.1. The characteristic values of length, stress and strain rate for all the four rock types are summarised in **Table 3**.

From section 3.2.1, of the many theoretical fragmentation models, the most relevant models for primary fragments are Grady (1982); Glenn and Chudnovsky (1986) and Zhou et al. (2006) models. In order to compare the experimental results with the existing theoretical models, the average fragment size needs to be appropriately normalized. The expression for normalised mean fragment size as per the theoretical model of Grady (1982), Glenn and Chudnovsky (1986) and Zhou et al. (2006) with the normalized strain rate are listed in **Eq. 15-17** (Levy and Molinari 2010):

$$\bar{S}_{Grady} = \left(\frac{24}{\bar{\varepsilon}^2}\right)^{\frac{1}{2}} \tag{15}$$

$$\bar{S}_{GC} = \frac{4}{\bar{\varepsilon}} \sinh\left(\frac{1}{3} \sinh^{-1}\left(\frac{3}{2}\bar{\varepsilon}\right)\right) \tag{16}$$

$$\bar{S}_{Zhou} = \frac{4.5}{1 + 4.5 \, \bar{\varepsilon}^{\,2/3}} \tag{17}$$

where, $\bar{\dot{\varepsilon}} = \frac{\dot{\varepsilon}}{\dot{\varepsilon}_o}$ ; $\bar{S} = \frac{S_o}{L_o}$

The fragment size results from the present study for primary and secondary fragments are summarised in **Fig. 11a.** Although power law relation might simplify, it is most commonly used in the study of fragment size. The power law fits very well to the experimental data, **Fig. 11a** shows that the normalised mean particle size of primary fragments gradually decreases with increase in the strain rate and remains flat at intermediate strain rate ($\dot{\varepsilon} > 10^1$) onwards. In the case of secondary fragments, the mean fragment size begins to flatten at lower strain rate ($10^0 < \dot{\varepsilon} < 10^1$) onwards.

The fragmentation results of the present study for mean particle size of primary fragments are compared with the theoretical models in the non-dimensional log-log plot in **Fig. 11b.** Although, none of these theoretical models predict the exact experimental fragment size, the trend of the experimental data is more similar to the Glenn and Chudnovsky (GC) model. However, the magnitude of the fragment size from experiments are three times lower than the GC model. Moreover, the strain rate sensitivity in GC models appears to begin at low strain rates ($10^0 < \dot{\varepsilon} < 10^1$), whereas in the present experiments, the fragments size begins to decrease at intermediate strain rate onwards ($\dot{\varepsilon} > 10^1$). A global power law relation defining the rate dependency of the mean particle size of primary ($\bar{S}_p$) and secondary ($\bar{S}_s$) fragments from the experiments are given as :

$$\bar{S}_p = 0.69 \pm 0.03 \ \bar{\dot{\varepsilon}}^{\ -0.019 \pm 0.031} ; \ \bar{S}_s = 0.125 \pm 0.01 \ \bar{\dot{\varepsilon}}^{\ -0.513 \pm 0.013} \tag{18}$$

No specific model is available for comparison of the secondary finer debris and the present experimental data cannot be directly compared with the existing theoretical models. But for the sake of completeness, the experimental results of secondary fragments are cautiously correlated in the same plot adjacent to primary fragments. The power law for secondary fragments appear to have a linear decreasing trend at low to intermediate strain rate. The secondary fragment sizes are significantly lower (~ an order of magnitude) than the primary fragment size. The power law for primary fragments of dynamic Brazilian tests is nearly entirely independent of strain rate. However, at intermediate strain rate, there are signs of decrease in the fragment size. Additional investigation at higher strain rate will determine if there is any significant effect of strain rates on the fragment size thereafter.

## 4  Summary and Conclusion

In this study, we report on 40 dynamic Brazilian experiments for estimating the tensile strength and fragment size at low to intermediate strain rate (in the range of $10^0$ to $2.7 \times 10^1$ /s). Four different rock lithologies are considered, of which two are igneous rocks (basalt and granite) and the other two are from sedimentary (sandstone) and metamorphic type (marble) respectively. We demonstrate that reliable strain rate measurements are possible using a centrally mounted strain gauge in the flattened Brazilian rock samples. The experimental results show that the split tensile strength of the rock is dependent on strain rate, with sudden increase in strength by a factor 2 is observed as the characteristic strain rate. The average characteristic strain rate in tension for basalt, granite, sandstone and marble are found to be 2.40 ± 0.68, 2.52 ± 1.0, 2.61 ± 0.56 and 2.39 ± 1.15, respectively. Moreover, the characteristic strain rate in tension is found to be approximately 1 to 2 orders of magnitude lower than the characteristic value of the same rock in compression. The split tensile strength of rocks in a unified form expressed in terms of characteristic strain rate and characteristic stress, has a rate of increase exponent factor of 0.583 ± 0.012. Considering the influence of rocks inhomogeneity and non-linear behaviour, the experimental results are very much in accordance with universal theoretical scaling model with exponent two-thirds, as predicted by Kimberley et al (2013).

The study showed that fragmentation in split tension mode will be vital in understanding various phenomena, where indirect tension failure and compression induced spallation failure take place. The fragment size distribution is determined for two class of fragments, namely, coarse sized primary fragments and finer secondary fragments. The mean fragment sizes of rocks in the primary and secondary assembly are described by a power law function of strain rate. The experimental results do not correspond to any of the existing theoretical models, but the mean particle size of primary fragment are found to be have a behaviour similar to Glenn and Chudnovsky's model at lower strain rates, where fragment size remains nearly constant up to the transitional strain rate, and decrease thereafter. It can be experimentally stated that the theoretical models is partially successful in predicting the dominant fragment size that fail in the dynamic split tension mode. With regard to secondary fragments, the finer fragment size appears to follows the linear decreasing trend in the log-log plot and the fragment size values are lower by an order magnitude compared to primary fragment size. In addition, it is important to note that the secondary fragments from the experiments are a major by-product and have significant role in tensile fragmentation, particularly at intermediate strain rate.

## Acknowledgements

The financial support provided by DFG (Deutsche Forschungsgemeinschaft) project DFG-SCHA1612 / 2-1 is gratefully acknowledged. The authors acknowledge the efforts of colleagues and non-technical staffs in the Dept. of Geology, University of Freiburg and Fraunhofer Institute for High-Speed Dynamics (EMI), Germany. In particular, the authors thank Herbert Ickler and Gordon Mette for sample preparation and Louis Müller and Matthias Dörfler during the experiments. We also appreciate the technical help of Sebastian Hess with SHPB and Mike Weber for helping with the installation of strain gauges.

**Table 1** Physical and Mechanical properties of the investigated rocks

| Parameters | Basalt (**HeBa**) | Granite (**MaGr**) | Sandstone (**SeSa**) | Marble (**CaMa**) |
|---|---|---|---|---|
| Density (g/cm$^3$), $\rho$ | 2.90 ± 0.01 | 2.62 ± 0.01 | 2.04 ± 0.02 | 2.70 ± 0.01 |
| Tensile strength (MPa), $\sigma_t$ | 15.55 ± 5.1 | 8.38 ± 2.5 | 4.39 ± 1.4 | 6.26 ± 0.8 |
| Elastic Modulus[#], (GPa), $E_s$ | 63.1 ± 7.3 | 36.1 ± 1.5 | 13.8 ± 0.8 | 44.8 ± 3.2 |
| Elastic wave speed* (km/s), c | 4.66 ± 0.27 | 3.71 ± 0.16 | 2.58 ± 0.17 | 4.07 ± 0.3 |

*# $E_s$ value for Granite are obtained from Rae et al. (2020) and for rest of the rocks the data are obtained from Ra et al (2020), pers. comm., 2 Oct; * c = sqrt(Es/ρ)*

**Table 2** Dynamic split tensile strength and strain rate results of sample tested

| Sample | Tensile Strength (MPa) | Strain rate (/s) | Sample | Tensile Strength (MPa) | Strain rate (/s) |
|---|---|---|---|---|---|
| HeBa01 | 79.8 ± 4.8 | 20.1 ± 1.1 | SeSa01 | 15.8 ± 1.8 | 7.4 ± 0.7 |
| HeBa02 | 66.7 ± 3.5 | 13.7 ± 0.3 | SeSa02 | 13.8 ± 2.0 | 6.1 ± 0.3 |
| HeBa03 | 47.9 ± 2.5 | 5.4 ± 0.9 | SeSa03 | 14.5 ± 1.6 | 4.7 ± 0.3 |
| HeBa04 | 18.9 ± 1.0 | 0.4 ± 0.5 | SeSa04 | 12.3 ± 1.7 | 5.5 ± 0.8 |
| HeBa05 | 35.4 ± 1.8 | 1.5 ± 0.2 | SeSa05 | 12.6 ± 0.7 | 2.8 ± 0.1 |
| HeBa06 | 63.7 ± 8.4 | 15.4 ± 0.4 | SeSa06 | 8.0 ± 1.6 | 1.6 ± 0.4 |
| HeBa07 | 71.0 ± 7.9 | 9.8 ± 0.8 | SeSa07 | 24.4 ± 7.1 | 20.9 ± 1.5 |
| HeBa08 | 73.8 ± 3.8 | 12.9 ± 0.6 | SeSa08 | 22.2 ± 4.9 | 16.6 ± 1.1 |
| HeBa09 | 52.2 ± 3.1 | 8.6 ± 0.8 | SeSa09 | 15.0 ± 4.1 | 9.0 ± 0.9 |
| HeBa10 | 27.9 ± 2.1 | 2.0 ± 0.4 | SeSa10 | 23.9 ± 2.4 | 20.8 ± 0.7 |
| MaGr01 | 17.3 ± 1.5 | 3.1 ± 0.9 | SeSa11 | 16.7 ± 0.9 | 8.0 ± 0.5 |
| MaGr02 | 44.4 ± 5.3 | 24.2 ± 1.5 | SeSa12 | 12.4 ± 2.1 | 5.4 ± 0.3 |
| MaGr03 | 48.5 ± 12 | 27.1 ± 2.1 | CaMa01 | 26.5 ± 3.8 | 12.2 ± 2.4 |
| MaGr04 | 33.1 ± 4.0 | 10.0 ± 0.7 | CaMa02 | 38.2 ± 3.9 | 20.2 ± 0.6 |
| MaGr05 | 42.5 ± 7.2 | 12.4 ± 0.5 | CaMa03 | 15.2 ± 4.1 | - |
| MaGr06 | 30.7 ± 2.1 | 9.4 ± 0.6 | CaMa04 | 21.9 ± 1.6 | 11.3 ± 1.0 |
| MaGr07 | 16.7 ± 4.1 | - | CaMa05 | 23.3 ± 7.8 | 12.1 ± 0.9 |
| MaGr08 | 19.1 ± 1.8 | 7.0 ± 0.9 | CaMa06 | 9.1 ± 2.8 | - |
| MaGr09 | 29.8 ± 1.5 | 13.2 ± 1.1 | CaMa07 | 24.1 ± 15.48 | 15.4 ± 0.9 |
| MaGr10 | 17.4 ± 2.3 | 2.7 ± 0.3 | CaMa08 | 11.9 ± 1.01 | 1.1 ± 0.4 |

'-' indicates the strain gauge signals were failed to capture

Table 3 The estimated characteristic parameters for different rocks

| Rock Type | Characteristic Length (mm) $L_o$ | Characteristic Tensile Stress (MPa) $\sigma_o$ | Characteristic Tensile Strain rate(/s) $\dot{\varepsilon}_o$ |
|---|---|---|---|
| Basalt (**HeBa**) | 27.77 ± 18.22 | 16.38 ± 2.23 | 2.40 ± 0.68 |
| Granite (**MaGr**) | 42.61 ± 25.43 | 8.46 ± 1.68 | 2.52 ± 1.01 |
| Sandstone (**SeSa**) | 42.12 ± 26.86 | 4.97 ± 0.50 | 2.61 ± 0.56 |
| Marble (**CaMa**) | 34.44 ± 8.81 | 6.55 ± 1.56 | 2.39 ± 1.15 |

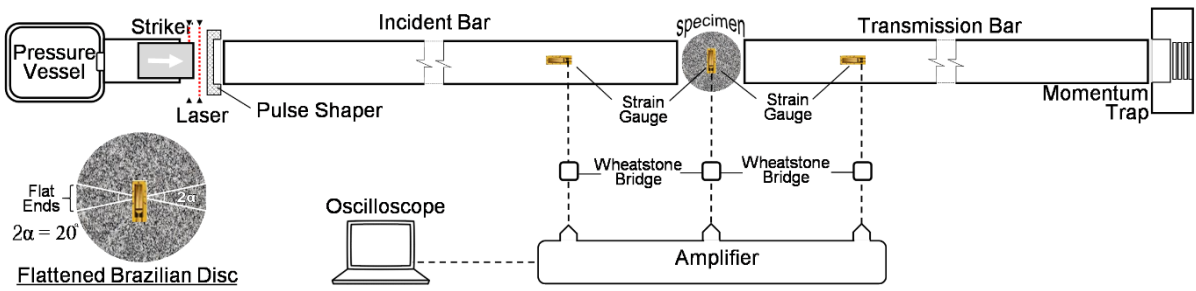

**Fig. 1** Schematic diagram of a Split Hopkinson Pressure bar and top view of the flattened Brazilian disc sample before mounting

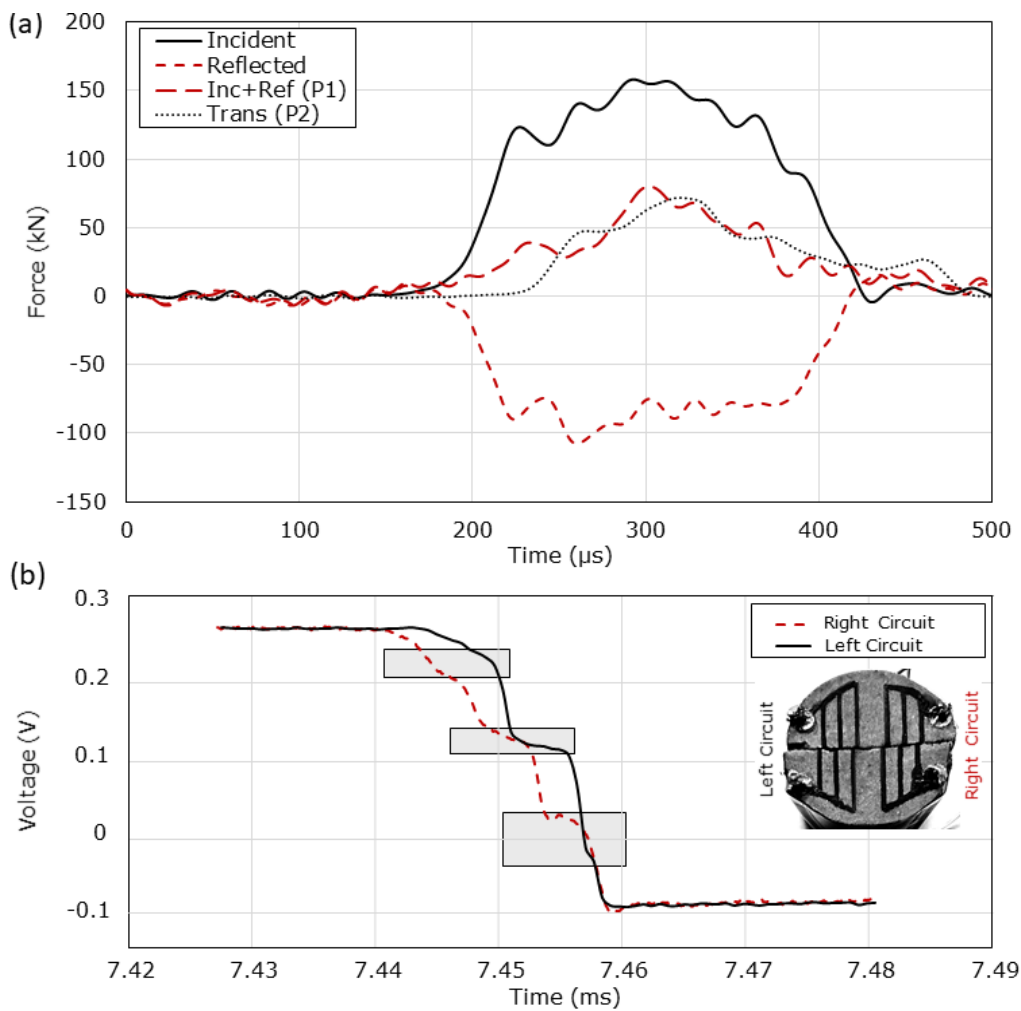

**Fig. 2 a** Dynamic forces evaluated at the ends of the sample, **b** Raw voltage signal recorded from the left and right electric paint circuit.

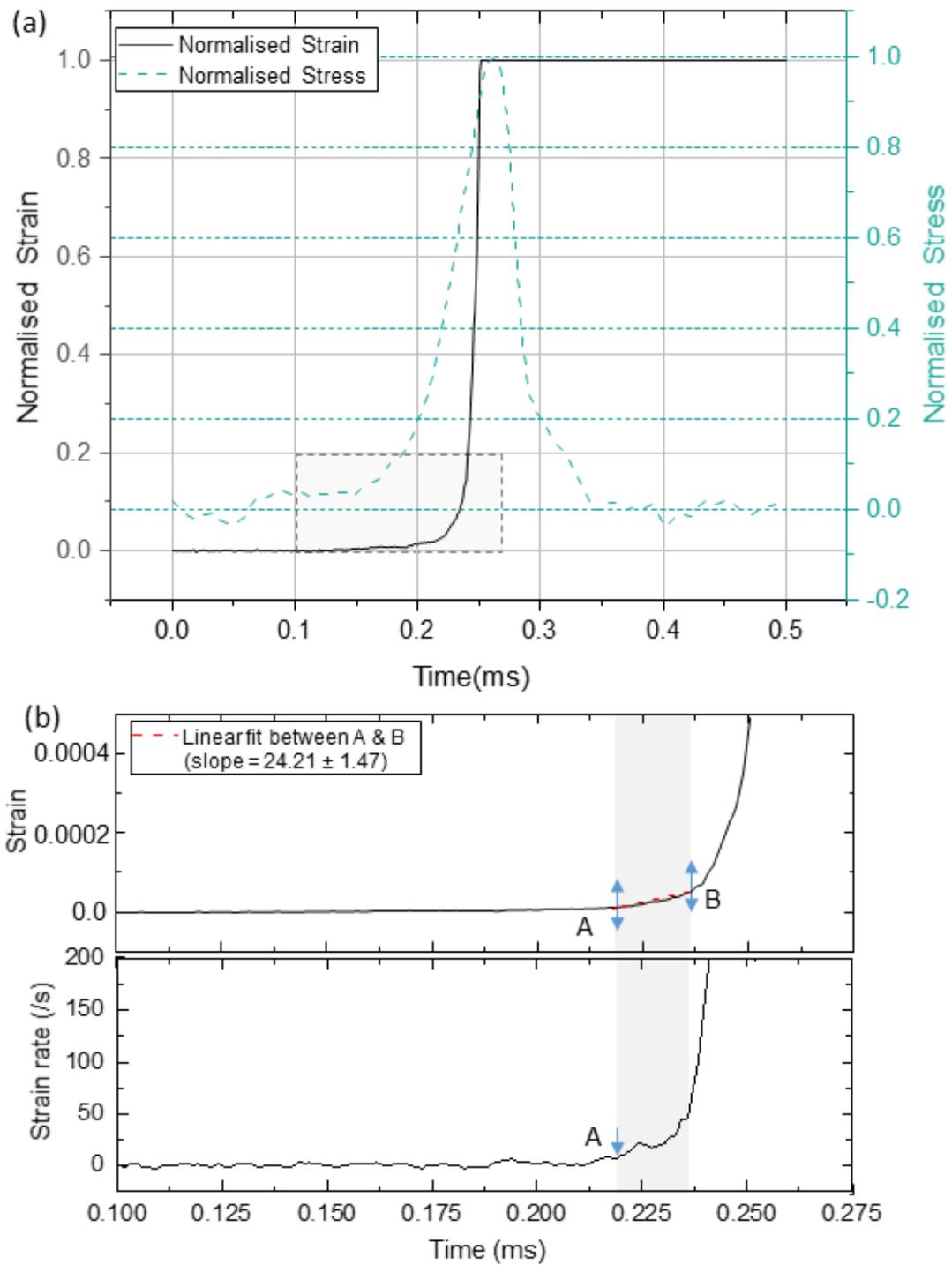

**Fig. 3 a** A typical plot of normalized strain and normalized stress versus time of a granite sample (MaGr02), **b** an extract of the strain and strain rate between 0.1 to 0.275 ms.

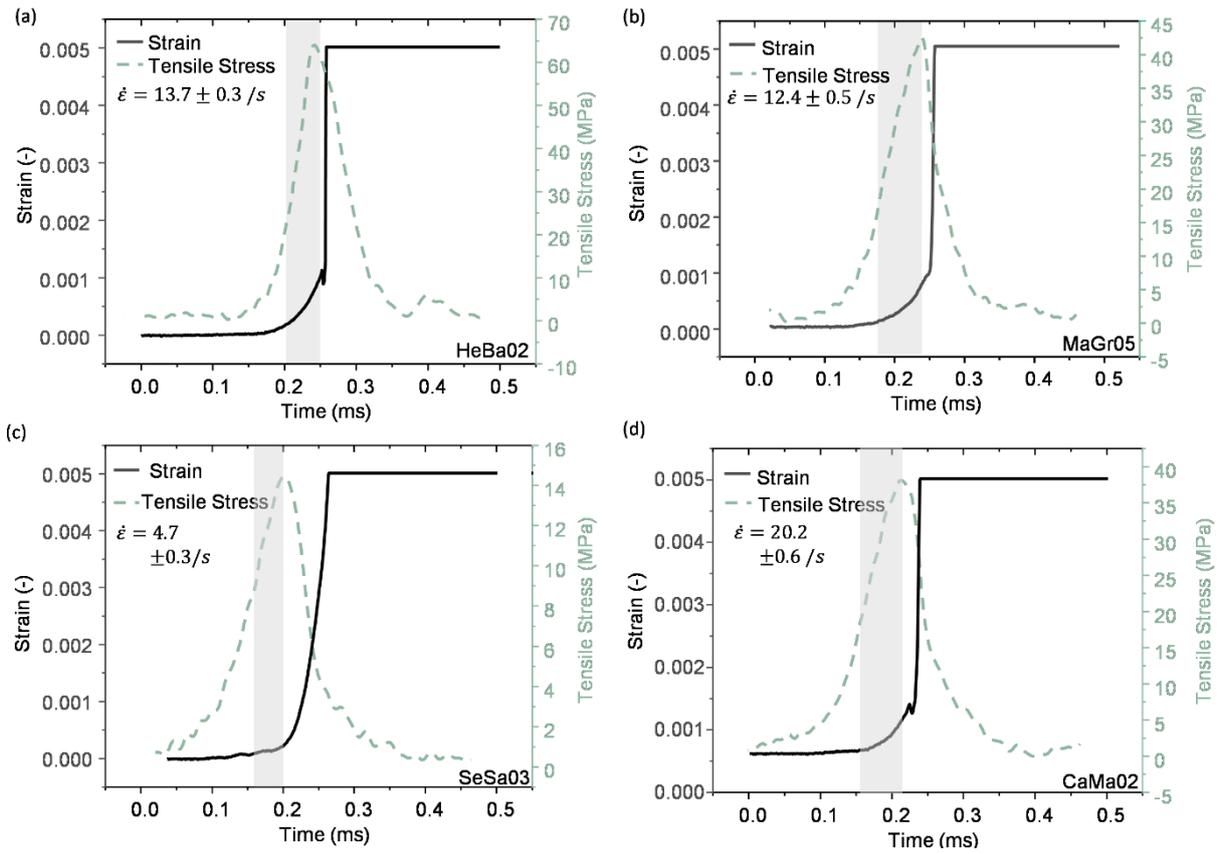

**Fig. 4** Tensile stress and strain against time for: **a** HeBa02, **b** MaGr05, **c** SeSa03 and **d** CaMa02

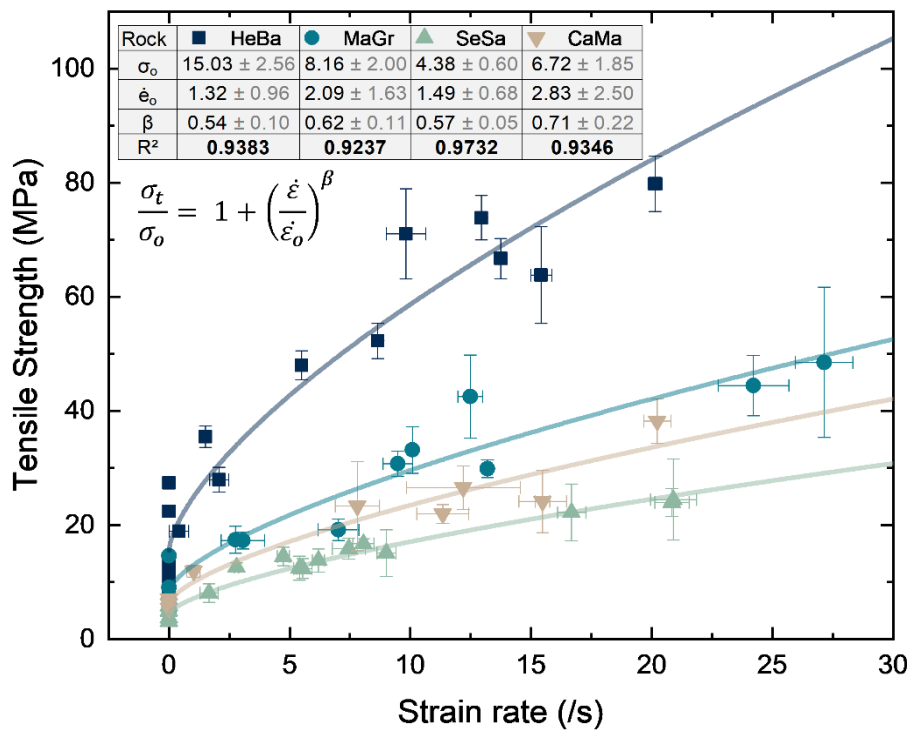

**Fig. 5** Variation of tensile strength with strain rate for the investigated rocks: Basalt, Granite, Sandstone and Marble

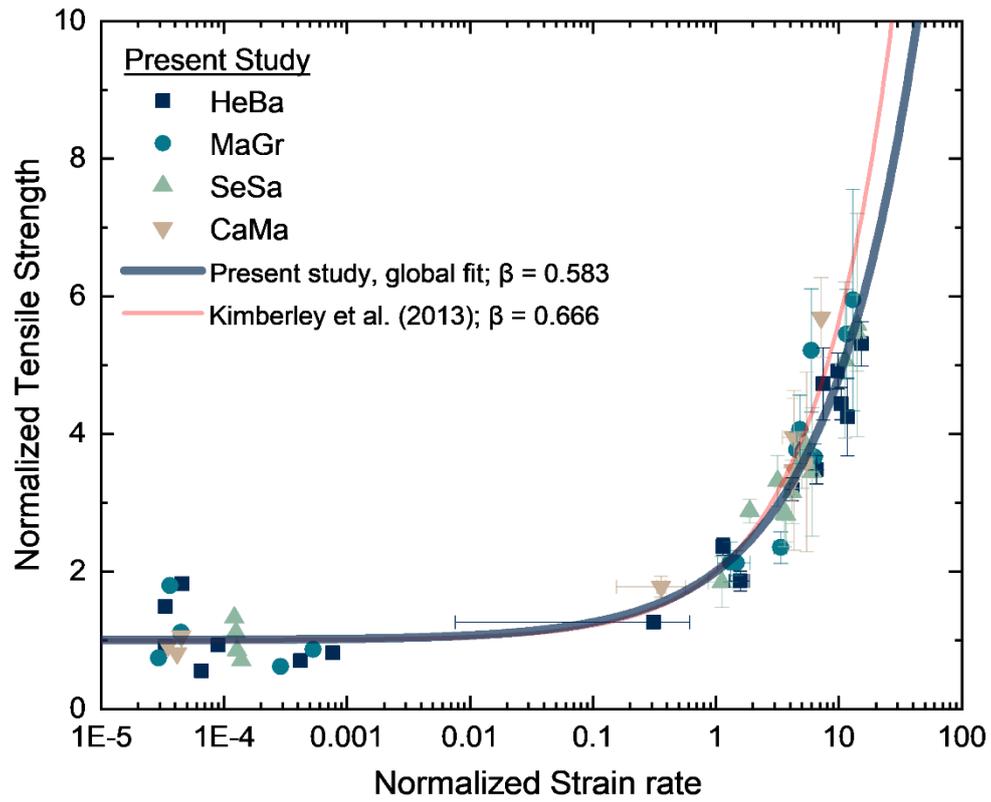

**Fig. 6** Normalized tensile strength data from the present experimental series is compared with the strength model of Kimberley et al. (2013)

**Fig. 7** Fragment morphology of Basalt (HeBa), Granite (MaGr), Sandstone (SeSa), Marble (CaMa) after dynamic Brazilian failure

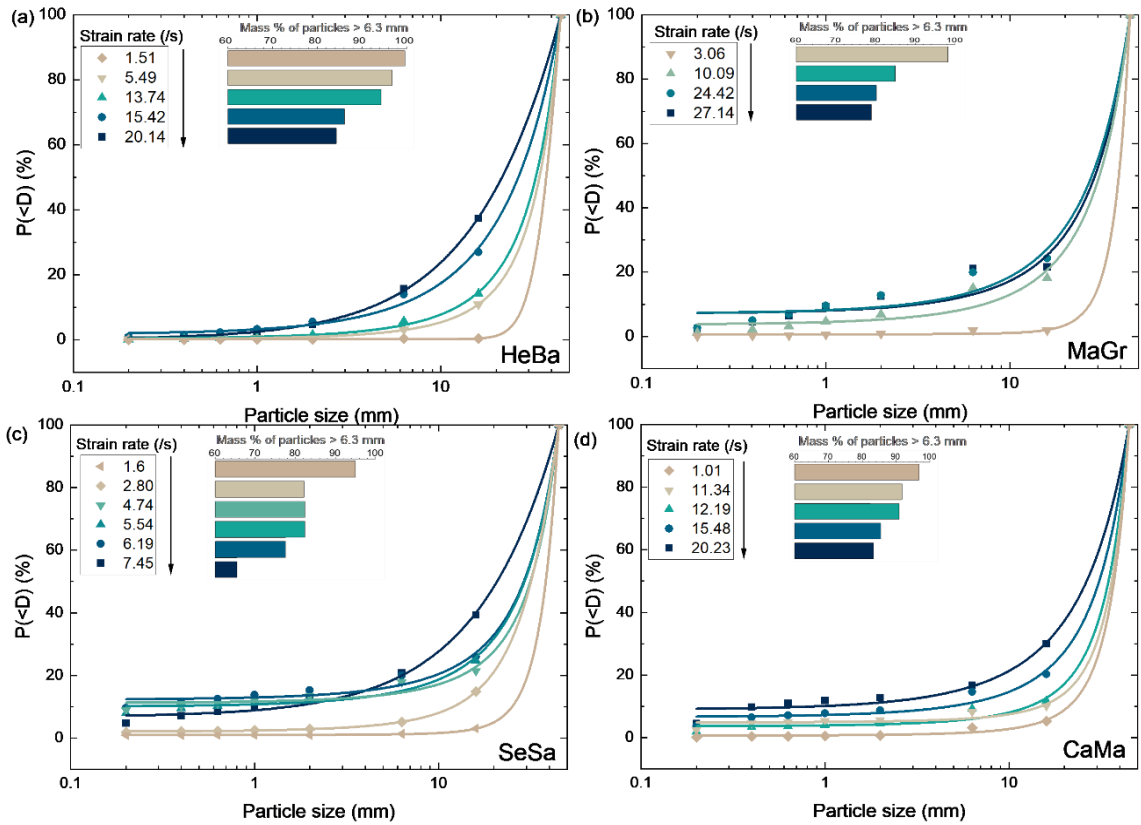

**Fig. 8** Fragment size distributions for the rocks: **a** Basalt, **b** Granite, **c** Sandstone and **d** Marble

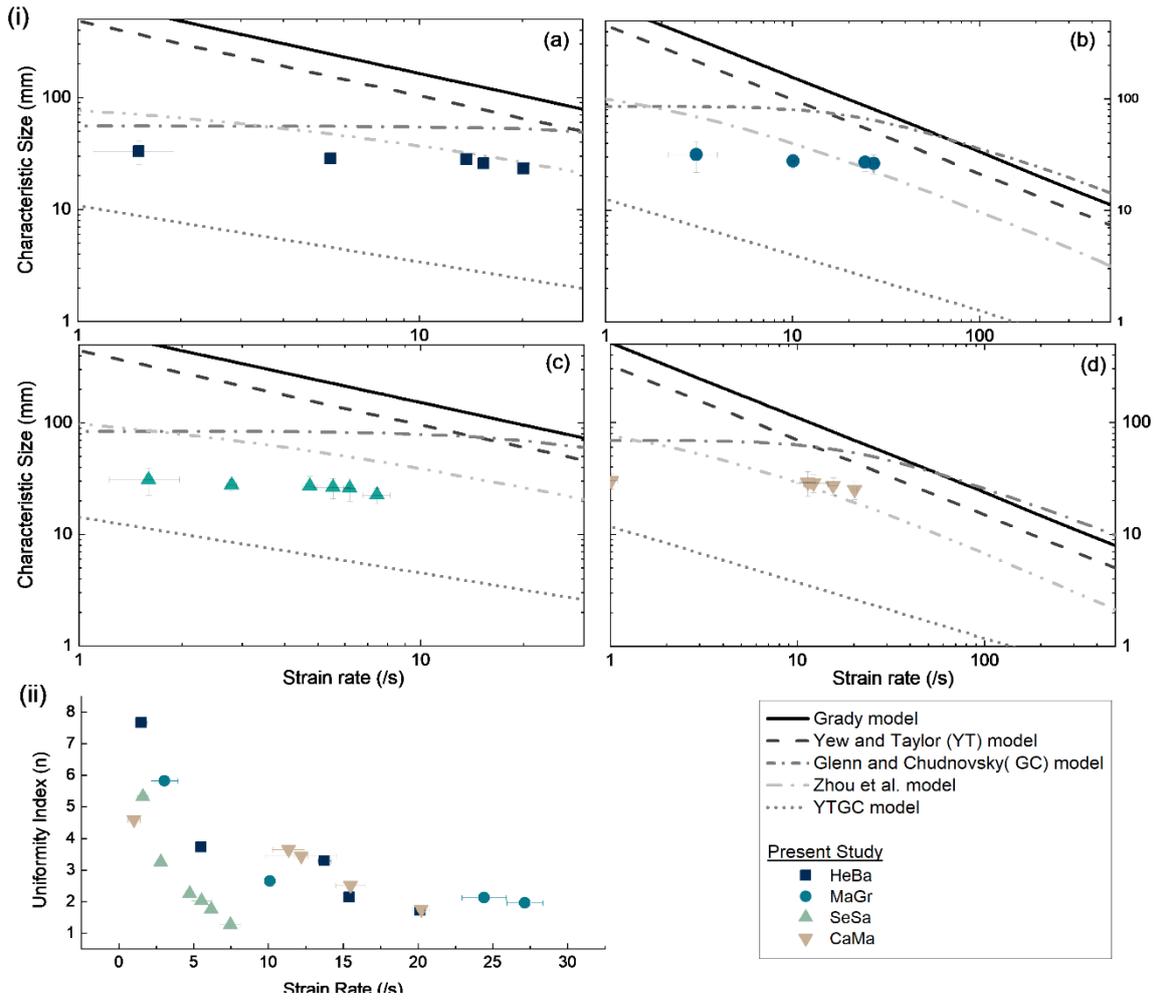

**Fig. 9 i** Dependency of characteristic fragment size on strain rate in rocks: **a** Basalt **b** Granite **c** Sandstone **d** Marble; **ii** Uniformity index as a function strain rates for primary fragments

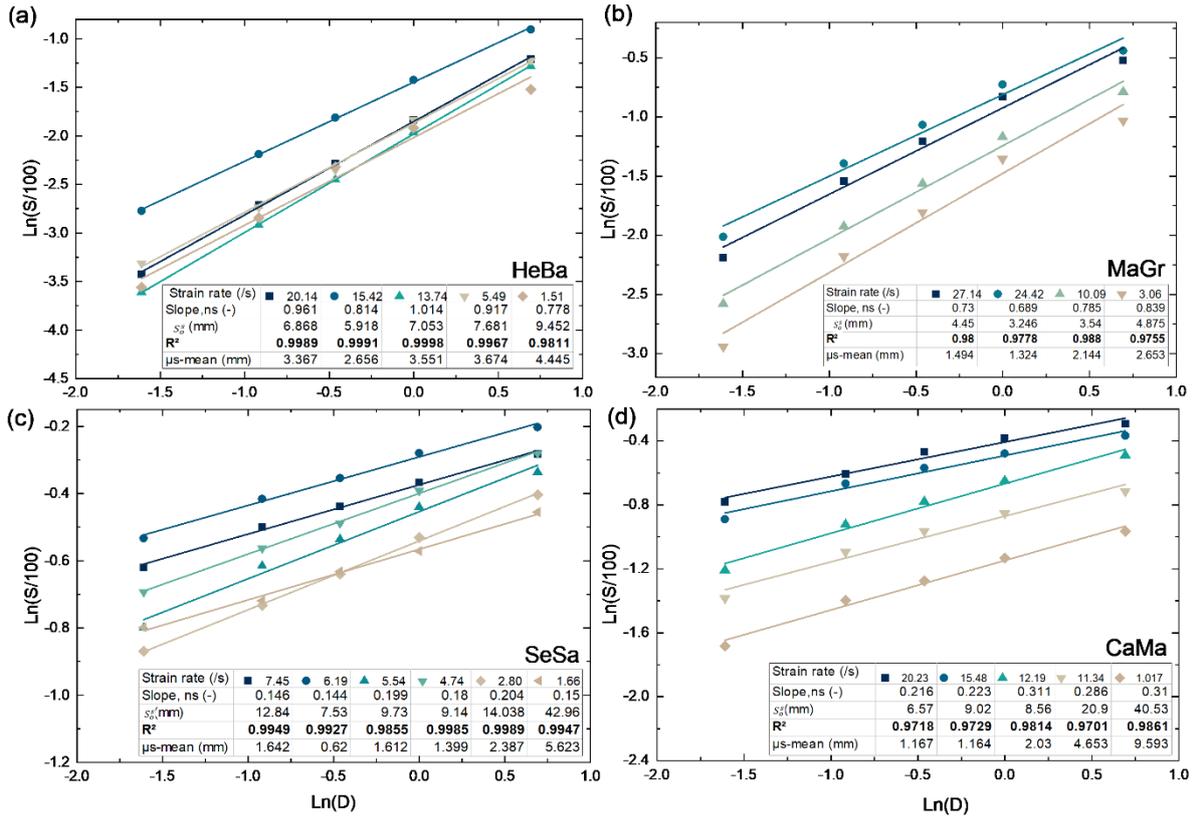

**Fig. 10** Grain size distribution of the secondary fracture debris for **a** Basalt, **b** Granite, **c** Sandstone and **d** Marble.

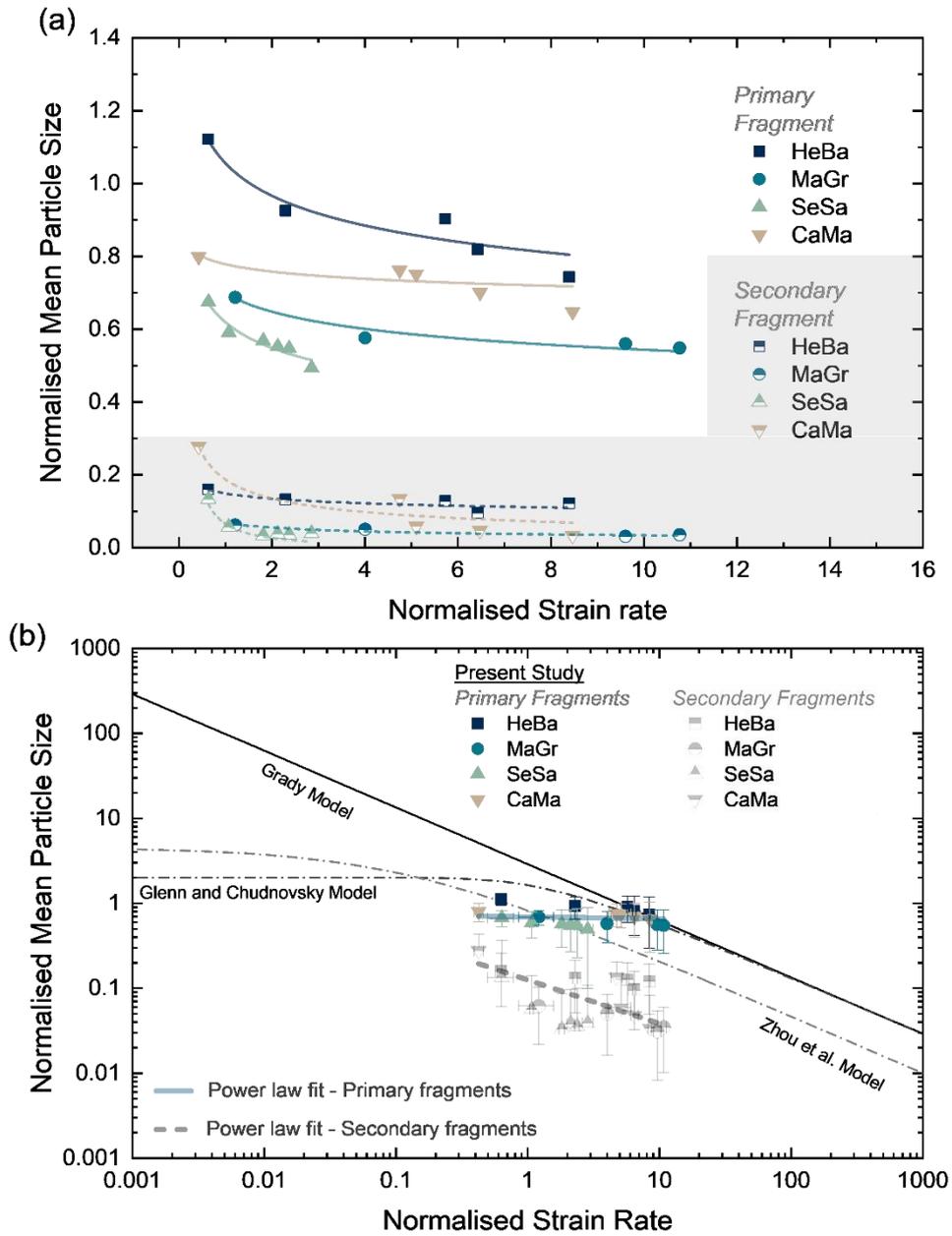

**Fig. 11** (a) An overview plot of normalised mean particle size (primary and secondary) versus normalised strain rate (b) A comparison of normalized mean particle size with different fragmentation models in log-log scale.